\documentstyle{article}

\begin{document}

\title{\bf Double Potts chain and exact results for some
two-dimensional models}

\author{ M.~A.~Yurishchev
\thanks{
On leave from Vasilsursk Laboratory,
Radiophysical Research Institute,
606263 Vasilsursk, Nizhny Novgorod Region, Russia
}
\thanks{E-mail: yur@itp.ac.ru}
}
\medskip
\date{\sl
The Abdus Salam International Centre for Theoretical Physics,
P.O.~Box 586, 34100 Trieste , Italy\\
}

\maketitle

\begin{abstract}
A closed-form exact analytical solution for the $q$-state Potts model
on a ladder $2\times\infty$ with arbitrary two-, three-, and four-site
interactions in a unit cell is presented.
Using the obtained solution it is shown that the finite-size
internal energy equation [6]
yields an accurate value of the critical temperature for the
triangular Potts lattice with three-site interactions in alternate
triangular faces.
It is argued that the above equation is exact at least for
self-dual models on isotropic lattices.
\end{abstract}

PACS: 05.50.+q, 05.70.Jk, 75.10.Hk

%----------------------------------------------------------------------

\newpage

\section{Introduction}
\label{sec:Intro}
The methods allowing to extract information about a
multi-dimensional system from solutions of their lower dimensional
counterparts play an important role in statistical physics.
One of the most well-known examples of this kind is
the finite-size scaling approach \cite{B83,P90}.

There are cases which evoke particular interest when the critical
properties of a system experiencing a phase transition can be
determined {\em exactly\/} from data only for its subsystems.
For instance, for the Ising strips the intersection point of the
locus of partition
function zeros in a complex temperature plane with the real
positive axis yields the exact value of the critical temperature
for the two-dimensional Ising model \cite{WTB87}.
Exact critical temperatures for the $S=1/2$ Ising models on
square, triangular, honeycomb, and centered square (Union Jack)
anisotropic lattices are obtained by using strip clusters when
an effective field is applied to one side of a strip only
\cite{LS92}.
Another exotic way of estimating the critical point of the square-lattice
Ising model was proposed in \cite{BS84}.
The authors of this paper showed that, in the quasi-diagonal
form of a transfer matrix of finite width strip, all
coefficients of the characteristic equation for the sub-block
containing the largest eigenvalue have extremum located precisely
at the exact value of phase transition temperature of infinite
lattice.

In the present paper we concentrate our attention on the method
to calculate the critical temperature, proposed by Wosiek \cite{W94}
(see also \cite{BW94,P95,BBSW95,SLS97,S98,S98a}).
The author \cite{W94} introduced  a maximum criterion
for the ratio of moments of the transfer matrix, and obtained the
following equation for determining the position of critical point in
a $d$-dimensional system:
\begin{equation}
   \label{eq:uu}
   u_1(K_c) = u_2(K_c) .
\end{equation}
Here $u_1$ and $u_2$ are the internal energies of $(d-1)$-dimensional
and two coupled $(d-1)$-dimensional subsystems, respectively, and
$K_c$ is the critical coupling (normalized inverse critical
temperature) of the $d$-dimensional system.

It is remarkable that at $d=2$ Eq.~(\ref{eq:uu}), shown in
\cite{W94}, yields the exact value of $K_c$ for the isotropic square
and triangular Ising lattices, as well as for the three-site Potts
model on the square lattice also with isotropic interactions.
Later on some other models were added to the list.
Now it includes another isotropic Baxter model (two square Ising
lattices coupled by four-particle interactions), the Baxter-Wu model
(triangular lattice with three-site interactions of Ising spins)
\cite{SLS97}, and $q$-state Potts model on the isotropic square
lattice with arbitrary value of $q$ \cite{S98a}.
Physical nature of Eq.~(\ref{eq:uu}) may be elucidated
when it yields either (a) an exact solution, or (b)
approximate estimation, or (c) does not give any solution at all
for a given model.
The more such examples are found, the better we understand its
sense.

For a two-dimensional system, Eq.~(\ref{eq:uu}) connects
the internal energies of infinitely long linear and double
chains.
Therefore, to obtain a possibility to test Eq.~(\ref{eq:uu}) rigorously,
it is necessary to have analytical solutions for such subsystems.

In the second section of the present paper, we give an exact
analytical solution for the two-chain Potts strip with a large
number of independent parameters.
As a special case, it contains a solution for the linear Potts chain.

Our solution for the double Potts chain enables us to cover all earlier
revealed cases in which Eq.~(\ref{eq:uu}) reproduces exactly the
critical temperatures for the two-dimensional models of Ising,
Baxter-Wu, and Potts.
Besides, we discover (section \ref{sec:TPL}) a new model for which
Eq.~(\ref{eq:uu}) yields the exact result.
This is the $q$-state Potts model on the triangular lattice with pure
three-site interactions in half of the triangular faces \cite{BTA78}.

In the fourth section of this paper a discussion of results is
presented.
In particular, it is shown that duality is a {\em sufficient\/}
condition for the validity of Eq.~(\ref{eq:uu}) for the isotropic
spin lattices.

Finally, in section 5, we summarize the results obtained in the work.

%---------------------------------------------------------------------

\section{Solution of double $q$-state Potts chain with
${\bf S}_q$ symmetry}
\label{sec:SDPC}

Consider a two-chain (ladder) lattice with spin variables
$\sigma_l^i$ attached to its sites ($i=1,2$ is the chain index,
and $l=1,2,3,\ldots$ labels the sites in the longitudinal direction
of the ladder); the spin variables take the values $1,2,\ldots,q$.

Let us write the Hamiltonian of a system in the form
\begin{equation}
   \label{eq:H}
   {\cal H} = -\sum_lH(\sigma_l^1,\sigma_l^2;
   \sigma_{l+1}^1,\sigma_{l+1}^2) .
\end{equation}
The locality of interactions in the above Hamiltonian allows one
to introduce the transfer matrix $V$ with elements
\begin{equation}
   \label{eq:V}
   \langle\sigma_1,\sigma_2|V|\sigma_1^{'},\sigma_2^{'}\rangle  =
   \exp[H(\sigma_1,\sigma_2; \sigma_1^{'},\sigma_2^{'})/k_BT]
\end{equation}
($T$ is the temperature and $k_B$ is Boltzmann's constant) and
reduce the problem of calculating the free energy density $f$ of
infinitely long strip to a search for the largest eigenvalue
$\lambda_1$ of the matrix $V$:
\begin{equation}
   \label{eq:f}
   f = \frac{1}{2}\ln\lambda_1 .
\end{equation}

The transfer matrix (\ref{eq:V}) has the size $q^2\times q^2$.
It is real and, moreover, all its elements are positive, but
the matrix is, generally speaking, non-symmetrical
($V_{ij}\neq V_{ji}$).

To solve the eigenvalue problem for the transfer matrix, we shall
use a group-theoretical approach (in this connection see, e.g.,
Ref.~\cite{Yu94} where such an approach was applied to a
quasidiagonalization of a transfer matrix of Ising model on
parallelepipeds $L\times L\times\infty$).
In order to obtain a solution for the two-leg spin ladder
(in which we are particularly interested) in a most general form,
we will go in reverse order.
Namely, we first select a symmetry group in the space
$|\sigma_1,\sigma_2\rangle$, that would enable us to quasidiagonalize
the transfer matrix up to sub-blocks secular equations which
can be solved analytically; only then we expand the
Hamiltonian density $H$ into a series on invariants of the
symmetry group.

Let us take a model which is invariant, for example, under
transformations of the symmetric group ${\bf S}_q$ of degree $q$.
For the Potts model this means, as a matter of fact, that we are
dealing with a system in the zero external field.
But fortunately the field is not required to test
Eq.~(\ref{eq:uu}).

It is known (see, for example, \cite{BN82}) that the largest eigenvalue of
a transfer matrix is located in the sub-block of the identity irreducible
representation (IR).
In accordance with group theory, the basis
vectors $\psi_i$ of the identity IR can be obtained  acting by
the permutation operators of a group ${\bf S}_q$ successively
on the orths $|1,1\rangle$, $|1,2\rangle$, $\ldots$\,,
$|q,q\rangle$.
Acting by elements of symmetric group first on the orth
$|1,1\rangle$ and then on the orth $|1,2\rangle$, we find that
the two linear combinations obtained involve all orths.
The normalized  basis vectors are given by
\begin{equation}
   \label{eq:psi}
   \psi_1 = \frac{1}{\sqrt{q}}\sum_{i=1}^q|i,i\rangle,\qquad
   \psi_2 = \frac{1}{\sqrt{q(q-1)}}\sum^q_{i,j=1}{}^{^{\!\prime}}
   |i,j\rangle
\end{equation}
(the prime at the second sum indicates that the terms with $i=j$ are
omitted).
Hence, the sub-block of identity IR has sizes 2 by 2
and, therefore, its eigenvalues (one of which is $\lambda_1$) can
easily be obtained by solving an algebraic equation only of second
degree.
Notice that if one takes the group ${\bf S}_q\times{\bf C}_s$
(${\bf C}_s$ is the group of mirror reflections in the plane going
between the chains of the two-leg ladder), the sub-block corresponding
to the identity IR will again have the size $2\times2$ and therefore
this symmetry, which only reduces the number of independent
parameters in a Hamiltonian, does not justify itself in the given
case.

We now represent the Hamiltonian (\ref{eq:H}) in the form of a sum
of terms which are invariant under transformations of the group
${\bf S}_q$:
\begin{eqnarray}
   \label{eq:HSq}
   & &
   {\cal H} = -\sum_l[
   J_1\delta_{\sigma^1_l\sigma^1_{l+1}} +
   J_2\delta_{\sigma^2_l\sigma^2_{l+1}} +
   J_0\delta_{\sigma^1_l\sigma^2_l} +
   J^{'}\delta_{\sigma^1_l\sigma^2_{l+1}} +
   J^{''}\delta_{\sigma^2_l\sigma^1_{l+1}}\nonumber\\
   & &\qquad
   + J_3\delta_{\sigma^1_l\sigma^2_l\sigma^1_{l+1}} +
   J_3^{'}\delta_{\sigma^1_l\sigma^1_{l+1}\sigma^2_{l+1}} +
   \tilde J_3\delta_{\sigma^1_l\sigma^2_l\sigma^2_{l+1}} +
   \tilde J_3^{'}\delta_{\sigma^2_l\sigma^1_{l+1}\sigma^2_{l+1}}\nonumber\\
   & &\qquad
   + J_4\delta_{\sigma^1_l\sigma^2_l\sigma^1_{l+1}\sigma^2_{l+1}}].
\end{eqnarray}
The Kronecker symbols entering here are defined as follows:
\begin{equation}
   \label{eq:Kr}
   \delta_{\sigma_1 \cdots\,\sigma_k} = \cases{
   1,&if $\sigma_1=\ldots =\sigma_k$\cr
   0,& otherwise} .
\end{equation}
The structure of two-site couplings in the Hamiltonian (\ref{eq:HSq})
is shown in Fig.~1.
Matrix elements of the original transfer matrix are written as
\begin{eqnarray}
   \label{eq:VSq}
   & &
   \langle\sigma_1,\sigma_2|V|\sigma_1^{'},\sigma_2^{'}\rangle  =
   \exp[
   K_1\delta_{\sigma_1\sigma_1^{'}} +
   K_2\delta_{\sigma_2\sigma_2^{'}} +
   {1\over 2}K_0(\delta_{\sigma_1\sigma_2} +
   \delta_{\sigma_1^{'}\sigma_2^{'}})\nonumber\\
   & &\qquad
   + K^{'}\delta_{\sigma_1\sigma_2^{'}} +
   K^{''}\delta_{\sigma_2\sigma_1^{'}} +
   K_3\delta_{\sigma_1\sigma_2}\delta_{\sigma_1\sigma_1^{'}} +
   K_3{'}\delta_{\sigma_1\sigma_1^{'}}\delta_{\sigma_1\sigma_2^{'}}\nonumber\\
   & &\qquad
   + \tilde K\delta_{\sigma_1\sigma_2}\delta_{\sigma_1\sigma_2^{'}} +
   \tilde K^{'}\delta_{\sigma_2\sigma_2^{'}}\delta_{\sigma_2\sigma_1^{'}} +
   K_4\delta_{\sigma_1\sigma_1^{'}}\delta_{\sigma_1\sigma_2}
   \delta_{\sigma_1\sigma_2^{'}}],
\end{eqnarray}
where $K_0=J_0/k_BT$, $K_1=J_1/k_BT$, $K_2=J_2/k_BT$, $K^{'}=J^{'}/k_BT$,
$K^{''}=J^{''}/k_BT$, $K_3=J_3/k_BT$, $K_3^{'}=J_3^{'}/k_BT$,
$\tilde K_3=\tilde J_3/k_BT$, $\tilde K_3^{'}=\tilde J_3^{'}/k_BT$, and
$K_4=J_4/k_BT$.

Using expressions (\ref{eq:psi}) and (\ref{eq:VSq}), we calculate
the matrix elements $Q_{ij}=\psi_i^{+}V\psi_j$ of the sub-block
corresponding to the identity IR:
\begin{eqnarray}
   \label{eq:Qij}
   & &
   Q_{11}=[q-1+\exp(K_1+K_2+K^{'}+K^{''}+K_3+K_3^{'}+\tilde K_3
   +\tilde K_3^{'}+K_4)]e^{K_0},\nonumber\\
   & &
   Q_{12}=(q-1)^{1/2}[q-2+\exp(K_1+K^{''}+K_3)+
   \exp(K_2+K^{'}+\tilde K_3)]e^{{1\over2}K_0},\nonumber\\
   & &
   Q_{21}=(q-1)^{1/2}[q-2+\exp(K_1+K^{'}+K_3^{'})+
   \exp(K_2+K^{''}+\tilde K_3^{'})]e^{{1\over2}K_0},\nonumber\\
   & &
   Q_{22}=(q-2)(q-3+e^{K_1}+e^{K_2}+e^{K^{'}}+e^{K^{''}})+
   \exp(K_1+K_2)+\exp(K^{'}+K^{''}).\nonumber\\
\end{eqnarray}
As a result, we find that the largest eigenvalue of the transfer
matrix of the double $q$-state Potts chain with the Hamiltonian
(\ref{eq:HSq}) reads
\begin{equation}
   \label{eq:L1}
   \lambda_1^{(2)} = {1\over2}(Q_{11}+Q_{22})+[{1\over4}(Q_{11}-
   Q_{22})^2+(q-1)Ae^{K_0}]^{1/2},
\end{equation}
where
\begin{eqnarray}
   \label{eq:A}
   & &
   A= [q-2+\exp(K_1+K^{''}+K_3)+ \exp(K_2+K^{'}+\tilde K_3)]
   [q-2\nonumber\\
   & &\qquad
   +\exp(K_1+K^{'}+K_3^{'})+
   \exp(K_2+K^{''}+\tilde K_3^{'})].
\end{eqnarray}

Those variants of the double Potts chains solved earlier
\cite{WTB87,S98a,S00,CS00,CS00a} correspond to a particular
choice of the interaction constants.
Setting $J_0=J^{'} (=J)$ with all other interaction
constants vanishing, we arrive at the solution for the linear
Potts chain \cite{P52}:
\begin{equation}
   \label{eq:L1D}
   \lambda_1^{(1)}(K) = e^K+q-1.
\end{equation}

%----------------------------------------------------------------------

\section{Triangular Potts lattice with three-site interactions on
alternate triangle faces}
\label{sec:TPL}

A large number of independent parameters in the model, solved in the
previous section, enables us to test Eq.~(\ref{eq:uu}) for a wide
class of two-dimensional spin systems.

In addition to the cases listed in the Introduction, in which
Eq.~(\ref{eq:uu}) is satisfied exactly, let us consider the Potts
model on a triangular lattice with three-site interactions in each
up-triangle (Fig.~2).
The position of the critical point in this model was found with both
three- and two-site interactions \cite{BTA78}.
However, it is known \cite{S98a} that for the triangle lattice with
pair couplings, Eq.~(\ref{eq:uu}) yields the exact result only for
the Ising case ($q=2$).
Therefore we discuss the model with purely three-site
interactions.
For this case,
\begin{equation}
   \label{eq:Kc}
   K_c = \ln(1+q).
\end{equation}
Let us show that this value satisfies Eq.~(\ref{eq:uu}) by subsystems in
the shape of strips with a periodic boundary condition in the
transverse direction.

The internal energy of one-dimensional subsystem is
\begin{equation}
   \label{eq:u1}
   u_1(K)\equiv\frac{\partial f_1}{\partial K} =
   [(q - 1)e^{-K} + 1]^{-1}.
\end{equation}
Substituting $K_3=\tilde K_3$ with all other interaction
constants vanishing, from (\ref{eq:f}), (\ref{eq:L1}), and (\ref{eq:A})
we obtain an expression for the free energy density of
double Potts chain:
\begin{eqnarray}
   \label{eq:f2}
   & &
   f_2(K) = {1\over 2}\ln[\![{1\over2}(e^{2K}+q^2-1)+
   [{1\over4}(e^{2K}-(q-1)^2)^2
   + q(q-1)(2e^K+\nonumber\\
   & &\qquad
   q-2)]^{1/2}]\!].
\end{eqnarray}
The internal energy $u_2(K)=\partial f_2/\partial K$.
Differentiating Eq.~(\ref{eq:f2}) with respect to $K$ one finds
an expression for $u_2(K)$.

An analysis shows that the dependences $u_1(K)$ and $u_2(K)$
have a crossing point which lies exactly at
$K=K_c=\ln(1+q)$ both for integer and non-integer $q$.
The internal energy of a system at critical point is
equal to
\begin{equation}
   \label{eq:uoo}
   u_{\infty}(K_c) = u_1(K_c) = u_2(K_c)
   = {1\over2}(1 + q^{-1}).
\end{equation}
Thus, using solutions only for the linear and double Potts chains,
Eq.~(\ref{eq:uu}) has enabled us to extract
the exact value of $K_c$ for the bulk two-dimensional Potts model
on a triangular lattice alternating faces which interact
by three-site forces.

%----------------------------------------------------------------------

\section{Discussion}

In the paper \cite{P95} Eq.~(\ref{eq:uu}) was extended up to
the form
\begin{equation}
   \label{eq:uLuL1}
   u_L(K_c) = u_{L'}(K_c)\qquad (L, L^{'} = 1, 2, 3,\,\ldots\,).
\end{equation}
Here $u_L$ is the internal energy per site of $L$ coupled
$(d-1)$-dimensional subsystems.
In the two-dimensional case $L$ denotes the width of a strip.

The validity of condition (\ref{eq:uLuL1}) for arbitrary $L$ and
$L^{'}$ means the absence of a ``singular'' (i.e. depending on $L$)
part of the internal energy density at critical point:
\begin{equation}
   \label{eq:uLKc}
   u_L(K_c) = const\quad {\rm upon}\quad L.
\end{equation}
In other words, the amplitudes of all finite-size corrections to the
critical internal energy of a system $u_{\infty}(K_c)$ are equal
to zero.

For the square isotropic Ising lattice, the derivative of the inverse
correlation length $\kappa_L(K)$ with respect to a temperature-like
variable $K$ has a similar property \cite{Yu00,Yu00a}:
\begin{equation}
   \label{eq:k1Kc}
   \frac{\partial\kappa_L}{\partial K}{\Biggr|}_{K=K_c} =
   \frac{\partial\kappa_{L^{'}}}{\partial K}{\Biggr|}_{K=K_c},
\end{equation}
i.e. $\partial\kappa_L/\partial K|_c$ does not depend on $L$.
This property has enabled us to determine exactly the value of
the thermal critical exponent $y_t\,(=1)$ for this model using only
the finite-size data \cite{Yu00,Yu00a}.

Equations (\ref{eq:uu}) and (\ref{eq:uLuL1}) are valid for the
{\em ferromagnetic\/} isotropic square Potts lattices.
These models are self-dual and the critical coupling for them
(in the anisotropic case) is determined from the condition
\begin{equation}
   \label{eq:Kcf}
   (e^{K_x} - 1)(e^{K_y} - 1) = q.
\end{equation}
For the {\em antiferromagnetic\/} square-lattice Potts model
the criticality condition is \cite{B82a}
\begin{equation}
   \label{eq:Kcaf}
   (e^{K_x} + 1)(e^{K_y} + 1) = 4 - q,
\end{equation}
where both $K_x < 0$ and $K_y < 0$.
We performed a verification and found that in the antiferromagnetic
case the curves $u_1(K)$ and $u_2(K)$ do not have any self-crossing
point and therefore Eq.~(\ref{eq:uu}) does not lead to the
exact value which follows from Eq.~(\ref{eq:Kcaf}) or even to any
approximate estimate for the critical point.

It is not difficult to show that if a model is self-dual and by this
the dual point coincides with the original one, then
Eqs.~(\ref{eq:uu}) and (\ref{eq:uLuL1}) are valid.

Indeed, consider for instance the Ising model on the isotropic
square lattice $L\times N$ with toroidal boundary conditions.
The partition function of such a system has a fundamental
property: it is invariant (up to multiplicative factor
exponentially depending on $L$) under the duality transformation
(see \cite{B82}):
\begin{equation}
   \label{eq:ZLN}
   Z_{L,N}(K^{*}) = (\sinh2K)^{-LM}Z_{L,M}(K),
\end{equation}
where $K$ and $K^{*}$ are related by the condition
\begin{equation}
   \label{eq:thK}
   \tanh K^{*} = e^{-2K}.
\end{equation}
(We here used another normalization of the exchanged constant in
the Ising model, namely $J_{\rm Potts}=2J_{\rm Ising}$.)
In the limit of an infinitely long strip ($N\to\infty$),
Eq.~(\ref{eq:ZLN}) transforms to the duality condition for the
largest eigenvalue:
\begin{equation}
   \label{eq:L1L}
   \lambda_1^{(L)}(K^{*}) = (\sinh2K)^{-L}\lambda_1^{(L)}(K).
\end{equation}
It follows from here that the values of the normalized internal
energy in dually conjugated points ($K$ and $K^*$) are connected
by the relation
\begin{equation}
   \label{eq:uLK}
   u_L(K^{*})\frac{\partial K^{*}}{\partial K} =
   u_L(K) - 2u_0(K)
\end{equation}
in which the additive term $u_0\ (=\coth 2K)$
{\em does not depend on\/} $L$.
Another important feature, this time related to isotropy of a
lattice, is that the dually conjugated points $K$ and $K^{*}$
confluent into one point at criticality:
\begin{equation}
   \label{eq:KK}
   K^{*} = K = K_c.
\end{equation}
Using condition (\ref{eq:thK}) we find that at the critical point
$K_c=(1/2)\ln(1+\sqrt2)$ the derivative
$\partial K^{*}/\partial K\bigr|_c=-1$.
Consequently
\begin{equation}
   \label{eq:uLu0Kc}
   u_L(K_c) = u_0(K_c) = \sqrt2.
\end{equation}
Thus, the critical internal energy per site $u_L(K_c)$ of
an Ising cylinder with isotropic square cells obeys the
condition (\ref{eq:uLKc}) for all $L=1,2,\ldots$.
That, in turn, leads to the validity of Eqs.~(\ref{eq:uu}) and
(\ref{eq:uLuL1}).

In an analogous way, Eqs.~(\ref{eq:uu}) and (\ref{eq:uLuL1})
can be derived for other isotropic spin models partition
functions which satisfy a functional equation like
\begin{equation}
   \label{eq:ZLg}
   Z_L(K^{*}) = [g(K)]^LZ_L(K).
\end{equation}

In those cases when the model is self-dual but the critical
manifold is a line or a surface (as, e.g., for anisotropic
lattices), Eqs.~(\ref{eq:uu}) and (\ref{eq:uLuL1}) no longer
hold.

This is not difficult to prove if we consider again a
two-dimensional Ising model.
For the anisotropic square lattice the duality condition
now looks as
\begin{equation}
   \label{eq:L1La}
   \lambda_1^{(L)}(K_x^{*}, K_y^{*}) =
   (\sinh2K_x\sinh2K_y)^{-L/2}\lambda_1^{(L)}(K_x, K_y)
\end{equation}
with
\begin{equation}
   \label{eq:thKxy}
   \tanh K_x^{*} = e^{-2K_y}\quad {\rm and}\quad
   \tanh K_y^{*} = e^{-2K_x}.
\end{equation}
From this it follows that on the critical line
\begin{equation}
   \label{eq:shKxy}
   \sinh2K_x \sinh2K_y = 1
\end{equation}
a condition (\ref{eq:L1La}) relates the values of the free
energy in {\em distinct\/} (dually conjugated) points
($K_x,K_y$) and ($K_y,K_x$):
\begin{equation}
   \label{eq:fLKxy}
   f_L(K_x, K_y) = f_L(K_y, K_x) + {1\over2}\ln(\sinh2K_x\sinh2K_y).
\end{equation}
This circumstance prevents fulfillment of the validity of
Eqs.~(\ref{eq:uu}) and (\ref{eq:uLuL1}) which identify
the internal energies at the same point.

The critical internal energy density of a strip $L\times\infty$
cut out from {\em anisotropic} lattice depends on the size $L$.
This is easy to verify if, using the results of the section 2,
one calculates the values $u_1(K_c)$ and $u_2(K_c)$ for the
anisotropic Ising and Potts lattices.

On the other hand, we can establish the same property if we take the
Onsager solution \cite{O44} for the two-dimensional Ising
model.
The dominant eigenvalue of a transfer matrix of cylinder
$L\times\infty$ with spatially anisotropic interactions
is equal to
\begin{equation}
   \label{eq:lam1}
   \lambda_1^{(L)}(K;\alpha) = (2\sinh 2K)^{L/2}\exp [(\gamma_1 +
   \gamma_3 + \ldots + \gamma_{2L-1})/2],
\end{equation}
where $\alpha=J_y/J_x$ is the lattice anisotropy parameter and
$\gamma_r$ are positive solutions of the equations
\begin{equation}
   \label{eq:chgam}
   \cosh\gamma_r = \cosh2\alpha K\coth 2K -
   \frac{\sinh2\alpha K}{\sinh2K} \cos\Bigg(\frac{\pi r}{L}\Bigg).
\end{equation}
From this, for the internal energy per site, we obtain
\begin{equation}
   \label{eq:uLKa}
   u_L(K;\alpha) = \coth2K + \frac{1}{2L}\Bigg(
   \frac{\partial\gamma_1}{\partial K} +
   \frac{\partial\gamma_3}{\partial K} +
   \ldots +
   \frac{\partial\gamma_{2L-1}}{\partial K}\Bigg).
\end{equation}
The dependences $\gamma_r(K)$ have a smooth extremum
(minimum) which lies in the isotropic case ($\alpha=1$)
exactly at $K=K_c$ and therefore
\begin{equation}
   \label{eq:gamK}
   \frac{\partial\gamma_r}{\partial K}\Bigg|_{K=K_c} = 0\qquad
   (r\neq0).
\end{equation}
As a result, the second term in Eq.~(\ref{eq:uLKa}) disappears
and the critical internal energy ceases to depend on $L$.
When $\alpha\neq1$ then Eq.~(\ref{eq:gamK}) does not have a
place and $u_L(K; \alpha)$ depends on the strip width in
a complicated way.
This explains the failures of exact calculations of $K_c$
from Eq.~(\ref{eq:uu})
in the anisotropic Ising lattice \cite{SLS97}.

Ending the section we note that, in spite of Eqs.~(\ref{eq:uu})
and (\ref{eq:uLuL1}), Eq.~(\ref{eq:k1Kc}) cannot be
deduced from a dual invariance of system.

%----------------------------------------------------------------------

\section{Conclusions}

Using a group-theoretical approach we obtained the
exact analytical solution for the double Potts chain with the
Hamiltonian (\ref{eq:HSq}).
The solution gives a possibility to examine Eq.~(\ref{eq:uu})
for a large number of models both with Ising ($q=2$) and arbitrary
Potts spins (including non-integer $q$).
The validity of Eq.~(\ref{eq:uu}) for the
triangular Potts lattice with pure three-site interactions in
alternate triangular faces was established.

In this paper it was also shown that Eqs.~(\ref{eq:uu}) and
(\ref{eq:uLuL1}) are a consequence of a dual symmetry
of models for which the critical point coincides with its
dual image.

As far as the author knows, the inverse theorem has not been
proved.
Duality plus isotropy or, more correctly, self-conjugation of a
critical point are not {\em necessary\/} conditions for
Eq.~(\ref{eq:uu}).
Therefore, generally speaking, there can exist systems which are
not invariant under dual transformation or a combination of
dual and star-triangular ones, but for which all amplitudes of
finite-size corrections to the critical internal energy (or to
some other quantity) are equal to zero.

%----------------------------------------------------------------------

\section*{Acknowledgements}

The author thanks A.~A.~Belavin, A.~A.~Nersesyan, and A.~M.~Sterlin
for useful discussions and comments.
I am also grateful to the Abdus Salam International Centre for Theoretical
Physics (Trieste, Italy) for kind hospitality where
this work was finished.
The research presented in this paper is supported in part by the
grants RFBR No.~99-02-16472 and CRDF No.~RP1-2254.

%----------------------------------------------------------------------

%----------------------------------------------------------------------

\newpage

\section*{Figure captions}\ \\

Fig.~1.\
Geometry of two-site couplings in the double $q$-state Potts chain
with ${\bf S}_q$ symmetry.\\

Fig.~2.\
Fragment of Potts lattice with three-site interactions in
alternate triangular faces (shaded).

%----------------------------------------------------------------------

\end{document}